%% file: a_DtoA0enu.tex
\documentclass[aps,prl,twocolumn,showpacs,byrevtex]{revtex4-1}
\usepackage{times}
\usepackage{amsmath}
\usepackage{amsfonts}
\usepackage{amssymb}
\usepackage{amsthm}
\usepackage{bm}
\usepackage{latexsym}
\usepackage{graphicx}
\usepackage{color}
\usepackage{url}
\usepackage{subfigure}
\usepackage{ulem}
\usepackage{siunitx}
\newcommand{\BR}{\mathcal{B}}

\newcommand{\dochannel}{D^0 \to a_0(980)^- e^+ \nu_e}
\newcommand{\dpchannel}{D^+ \to a_0(980)^0 e^+ \nu_e}
\newcommand{\aomdecay}{a_0(980)^- \to \eta \pi^-}
\newcommand{\aoodecay}{a_0(980)^0 \to \eta \pi^0}
\newcommand{\doresultnsig}{25.7_{-5.7}^{+6.4}}
\newcommand{\dpresultnsig}{10.2_{-4.1}^{+5.0}}
\newcommand{\doresulta}{(1.33_{-0.29}^{+0.33}({\rm stat})\pm0.09({\rm syst}))\times10^{-4}}
\newcommand{\dpresulta}{(1.66_{-0.66}^{+0.81}({\rm stat})\pm0.11({\rm syst}))\times10^{-4}}
\newcommand{\doresult}{(1.33_{-0.29}^{+0.33}\pm0.09)\times10^{-4}}
\newcommand{\dpresult}{(1.66_{-0.66}^{+0.81}\pm0.11)\times10^{-4}}
\newcommand{\dpul}{<3.0\times10^{-4}}
\newcommand{\dpupnsig}{18.5}
\newcommand{\dosig}{6.5\sigma}
\newcommand{\dpsig}{3.0\sigma}
\newcommand{\dosigs}{6.4\sigma}
\newcommand{\dpsigs}{2.9\sigma}

\begin{document}

\title{\boldmath Observation of the Semileptonic Decay $\dochannel~$ and Evidence for $\dpchannel~$ }

\input{authors}
\affiliation{}

\date{\today}

\vspace{0.4cm}
\begin{abstract}
    Using an $e^+e^-$ collision data sample of 2.93 fb$^{-1}$ collected at a center-of-mass energy
    of 3.773 GeV by the BESIII detector at BEPCII, we report the observation
    of $\dochannel$ and evidence for $\dpchannel~$ with significances of $\dosigs$
    and $\dpsigs$, respectively. The absolute branching fractions are determined
    to be $\mathcal{B}(\dochannel)\times\mathcal{B}(\aomdecay) = \doresulta~$
    and $\mathcal{B}(\dpchannel)\times\mathcal{B}(\aoodecay) = \dpresulta~$.
    An upper limit of $\mathcal{B}(\dpchannel)\times\mathcal{B}(\aoodecay)\dpul~$ is also
    determined at the $90\%~$ confidence level.
\end{abstract}


\maketitle
The $a_0(980)$, a scalar meson with isospin $I=1$, though well-established
experimentally, has a very intriguing internal structure. It is often
interpreted as a multiquark state or a $K\bar{K}$ bound state~\cite{Amsler:2013}.
There are  many  studies on the production of scalar mesons in the decays
of $D$ mesons~\cite{Oset:2015,Xiao-Dong:2017,Achasov:2012,Ke:2009,Bennich:2009}.
Semileptonic $D$ decays provide a pristine environment where the $a_0(980)$
is produced by an isovector combination of $u$ or $d$ and their anti-quark partners.
Therefore, it is an ideal test-bed to study the underlying structure,
due to its clear production mechanism and limited final state interactions.
The authors of Ref.~\cite{Lv:2010} conclude that the ratio ($R$) of the sum of
the branching fractions of $D^+ \to f_0(980) e^+ \nu$ and $D^+ \to \sigma e^+ \nu$
to that of $D^+ \to a_0(980) e^+\nu$ provides a model independent way
to distinguish the quark components of the light scalar mesons.
For example, if the ratio $R$ were close to 1, the four-quark picture is likely
to be ruled out, but if $R$ were close to 3, the two-quark picture is likely to be excluded.
However, due to the lack of statistics and high backgrounds, measurements of
these interesting decays have not been reported yet.

In this Letter, we present the first observation of the semileptonic decay
$\dochannel$ and evidence for $\dpchannel$. The data sample used in this
analysis was collected at center-of-mass energy $\sqrt{s}=3.773$ GeV
(near the nominal mass of the $\psi(3770)$)~by the BESIII detector at the
BEPCII collider, and corresponds to an integrated luminosity of 2.93\,fb$^{-1}$~\cite{BesLum}.

The BESIII detector is described in detail elsewhere~\cite{bes3}.
The detector has a geometrical acceptance of $93\%$ of $4\pi$.
It includes a multi-layer drift chamber (MDC) for measuring the
momenta and specific ionization energy loss ($dE/dx$) of charged
particles, a time-of-flight system (TOF) which contributes to charged particle
identification (PID), a CsI(Tl) electromagnetic calorimeter (EMC)
for detecting electromagnetic showers, and a muon chamber system
designed for muon identification.

A detailed GEANT4-based~\cite{geant4} Monte Carlo (MC) simulation of the
BESIII detector is used to determine the detection efficiencies and evaluate
the possible background sources. Events are generated by the generator
{\sc kkmc}~\cite{kkmc} using {\sc evtgen}~\cite{evtgen}, with the effects
of beam energy spread and initial-state radiation (ISR) being taken
into account.  Final-state radiation is treated via the {\sc photos}
package~\cite{photons}.

A double-tag analysis technique~\cite{MarkIII:1986} is employed; this takes advantage
of $D$ mesons produced via exclusive $D\bar{D}$ pair-production in the decay
of the $\psi(3770)$ resonance.
We reconstruct $\bar{D}$ mesons using specific hadronic decays,
producing a sample of single tag (ST) events.
We then search these ST events for the partner $D$ meson undergoing
the decay process of interest; successful searches result in our sample of
double tag (DT) events.  This strategy suppresses non-$D\bar{D}$
background effectively and provides a measurement of absolute branching
fractions independent of the integrated luminosity and the
$D\bar{D}$ production cross section. These absolute branching fractions
are calculated as
\begin{equation}\label{forAbsBR}
\BR_{\rm sig} =
   \frac{
         N_{\rm sig}^{\rm obs}
        }
        {
          \sum_{\alpha}N_{\rm tag}^{\rm obs,\alpha}
          \epsilon_{\rm tag,sig}^{\alpha}
         /\epsilon_{\rm tag}^{\alpha}
        }{\rm ,}
\end{equation}
in which $\alpha$ denotes the different ST modes, $N_{\rm tag}^{\rm obs,\alpha}$
is the ST yield for tag mode $\alpha$, $N_{\rm sig}^{\rm obs}$ is the sum of
the DT yields from all ST modes, and $\epsilon_{\rm tag}^{\alpha}$
and $\epsilon_{\rm tag,sig}^{\alpha}$ refer to the corresponding ST efficiency and the DT
efficiency for the ST mode $\alpha$ determined by MC simulations. In this approach,
most of the systematic uncertainties arising from the ST reconstruction are canceled.

The ST $\bar{D}$ mesons are reconstructed with the following final states:
$\bar{D}^0 \to K^+ \pi^-$, $K^+ \pi^- \pi^0$, $K^+ \pi^- \pi^+ \pi^-$, and
$D^- \to K^+ \pi^- \pi^-$, $K^+ \pi^- \pi^- \pi^0$, $K_S^0 \pi^-$, $K_S^0 \pi^- \pi^0$,
$K_S^0 \pi^+ \pi^- \pi^-$, $K^+ K^- \pi^-$. The charged particles $K^{\pm}$ and
$\pi^{\pm}$, as well as the neutral particles $\pi^0$ and $K_S^0$, are selected with
the same criteria as those in Ref.~\cite{ycp}. For the $\bar{D}^0 \to K^+ \pi^-$ final state,
requirements on the opening angle and the difference of the time of flight
of the two charged tracks are applied to reduce backgrounds from cosmic rays, Bhabha and
di-muon events \cite{d0semilep}.~Throughout this Letter, charge-conjugate
modes are implied, unless otherwise noted.

\begin{figure}
\begin{center}
\includegraphics[height=6cm]{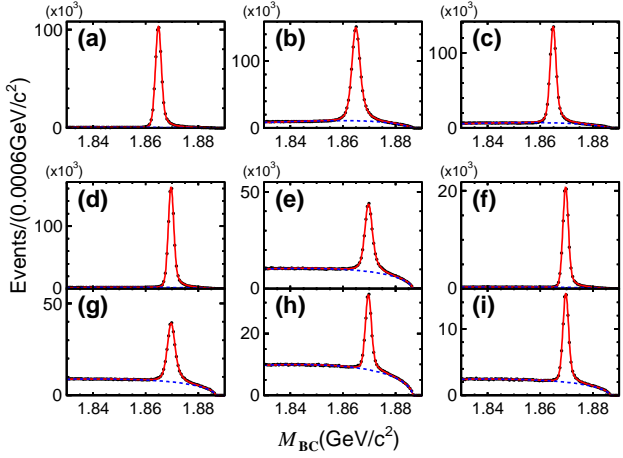}
\caption{(Color online). Fits to the $M_{\rm BC}$ distributions of
    the ST candidates. The first row shows the $\bar{D}^0$ modes:
    (a)~$K^+\pi^-$, (b)~$K^+\pi^-\pi^0$, (c)~$K^+\pi^-\pi^+\pi^-$,
    and the last two rows show the $D^-$ modes: (d)~$K^+\pi^-\pi^-$,
    (e)~$K^+\pi^-\pi^-\pi^0$, (f)~$K_S^0\pi^-$, (g)~$K_S^0\pi^-\pi^0$,
    (h)~$K_S^0\pi^+\pi^-\pi^-$, (i)~$K^+K^-\pi^-$.
    Points with error bars represent data, the (red) solid
    lines are the total fits and the (blue) dashed lines represent the
    background contributions.}
\label{fig:singleTagSample}
\end{center}
\end{figure}

Two key kinematic variables, the energy difference $\Delta E \equiv E_{D} - E_{\text{beam}}$,
and beam-constrained mass $M_{\rm BC} \equiv \sqrt{E_{\text{beam}}^{2}/c^{4} - |\vec{p}_{D}|^{2}/c^{2}}$,
are used to identify the ST $\bar{D}$ candidates.
Here, $E_{\text{beam}}$ is the beam
energy, $E_{D}$ and $\vec{p}_{D}$ are the reconstructed energy and
momentum of the $\bar{D}$ candidate in the $e^+e^-$ center-of-mass system.
For true $\bar{D}$ candidates, $\Delta E$ and $M_{\rm BC}$ will peak at
zero and the nominal mass of the $D$ meson, respectively. We accept the $\bar{D}$
candidates with $M_{\rm BC}$ greater than $1.83 ~{\rm GeV}/c^2$ and apply mode-dependent
$\Delta E$ requirements of approximately three standard deviations.
When multiple candidates exist, at most one candidate per tag mode per charm
(i.e., $D$ or $\bar{D}$) is retained in each event by selecting the candidate
with the smallest $|\Delta E|$~\cite{He:2005bs}.
The ST yields are determined by performing a maximum likelihood fit to the
$M_{\rm BC}$~distributions of the accepted $\bar{D}$ candidates, as shown
in Fig.~\ref{fig:singleTagSample}.  The signal shape is modeled by
the MC simulated shape convolved with a Gaussian function with free
parameters.  The MC simulation includes the effects of beam energy spread,
ISR, the $\psi(3770)$ line shape, and experimental resolution, while the
Gaussian allows for small imperfections in the MC simulation.
The combinatorial background is modeled
by an ARGUS function~\cite{Albrecht:1990am}.
The ST yield for each mode is calculated by subtracting the integrated
ARGUS background yield from the total number of events contained in
the signal regions defined as
$1.858 < M_{\rm BC} < 1.874 ~{\rm GeV}/c^2~$ for $\bar{D}^0$ and
$1.860 < M_{\rm BC} < 1.880 ~{\rm GeV}/c^2~$ for $D^-$.
The ST yields in data and the corresponding ST efficiencies are listed
in Table~\ref{tab:tagEfficiency}.

We search in the selected ST events for the semi-leptonic decays $\dochannel~$
and $\dpchannel~$, using the remaining charged tracks and photon
candidates not used for the ST candidate.
Here, the $a_0(980)^-$ and $a_0(980)^0$ are reconstructed by their
prominent decays to $\eta\pi^-$ and $\eta\pi^0$, respectively.
The PID of the charged hadrons (positrons) is accomplished by combining the
$dE/dx$ and TOF ($dE/dx$, TOF and EMC) information to construct a likelihood
$\mathcal{L}_i$ ($\mathcal{L}_i'$) for each of the hypotheses $i=e/\pi/K$.
The charged pion
candidate is required to satisfy $\mathcal{L}_{\pi} > \mathcal{L}_K$ and
$\mathcal{L}_{\pi}>0.1\%$. The positron candidate is required to satisfy
$\frac{\mathcal{L}_{e}'}{\mathcal{L}_{e}'+\mathcal{L}_{\pi}'\mathcal+{L}_{K}'}>0.8$
and $E/(pc) > 0.8$, where $E$ is the energy deposited in the EMC and $p$ is
the momentum measured by the MDC.  A candidate signal event is required
to have a single positron (electron) for signal $D$ ($\bar{D}$) decays.
The $\pi^0$ and $\eta$ candidates are formed from
pairs of photon candidates with invariant two-photon masses within $(0.115, 0.150)~$ and
$(0.508, 0.572)~{\rm GeV}/c^2$, respectively. To improve the kinematic resolution,
a one-constraint (1-C) kinematic fit is performed by constraining the $\gamma\gamma$
invariant mass to the expected nominal mass~\cite{pdg:2016}.  Background consisting
of a real photon paired with a fake one is suppressed by requiring the decay angle, defined
as $|\cos\theta_{{\rm decay},\pi^0(\eta)}| = \frac{|E_{\gamma1} - E_{\gamma2}|}{|\vec{p}_{\pi^0(\eta)}c|}$,
to be less than $0.80$ and $0.95$ for the $\pi^0$ and $\eta$ candidates,
respectively. Here, $E_{\gamma1}$ and $E_{\gamma2}$ are the energies of the two
daughter photons of the $\pi^0$($\eta$), and $\vec{p}_{\pi^0(\eta)}$ is the
reconstructed momentum of the $\pi^0$($\eta$). The $a_0(980)^-$ candidate is
formed with a charged pion and a selected $\eta$ candidate. The $a_0(980)^0$
candidate is formed from the combination of $\pi^0$ and $\eta$ candidates
with the least $\chi_{{\rm 1C}, \pi^0}^2 + \chi_{{\rm1C},\eta}^2$, where
$\chi_{{\rm 1C}, \pi^0}^2$ and $\chi_{{\rm 1C}, \eta}^2$ are the $\chi^2$
values of the 1-C kinematic fits of the $\pi^0$ and $\eta$ candidates,
respectively.  After the above selections, we veto any event with extra
unused charged tracks.
Events containing an additional unused $\pi^0$ candidate are
also rejected.
This $\pi^0$ veto suppresses the following backgrounds:
$D^0 \to \rho^- e^+ \nu_e$ and $D^0 \to K^*(892)^-e^+ \nu_e$
(with ($K^*(892)^-\to K_S^0\pi^-$)
for the $\dochannel$ mode,
and $D^+ \to K_S^0 e^+ \nu_e$ and
$D^+ \to \bar{K}^*(892)^0e^+ \nu_e$
(with $\bar{K}^*(892)^0 \to K_S^0\pi^0$)
for $\dpchannel$; in all cases here, $K_S^0 \to \pi^0\pi^0$.
Detailed MC studies show that $D^0 \to K^*(892)^- e^+ \nu_e$
and $D^+ \to \bar{K}^*(892)^0 e^+ \nu_e$ followed by $\bar{K}^* \to K_L^0 \pi$
are prominent backgrounds, where the $K_L^0$ signal in the EMC can mimic the higher-energy
daughter of the $\eta$ candidate.
To suppress these background, the lateral moment~\cite{latmom} of EMC showers,
which peaks around $0.15$ for real photons but varies from $0$ to $0.85$
for $K_L^0$ candidates, is required to be within $(0, 0.35)$ for the higher-energy photon from the $\eta$ decay.  This requirement suppresses about
$70\%$ of the $K_L^0$ backgrounds, while retaining $95\%$ of the signal.

\begin{table}[hbtp]
\caption{{\small{ST yields in data, $N_{\rm tag}^{\rm obs}$, ST efficiencies,
    $\epsilon_{\rm tag}$, and DT efficiencies, $\epsilon_{\rm tag,sig}$,
    with statistical uncertainties, for each mode $\alpha$.
    Branching fractions of $K_S^0 \to \pi^+\pi^-$,
    $\pi^0 \to \gamma\gamma$ and $\eta \to \gamma\gamma$ are not included in the
    efficiencies. The first three rows are for $\bar{D}^0$ candidates and the
    last six rows are for $D^-$ candidates.}}}
\label{tab:tagEfficiency}
\begin{center}
\begin{footnotesize}
\renewcommand{\arraystretch}{1.0}
\scalebox{1}{
\begin{tabular}{cS[table-format=8.0(4)] cc}
    \hline\hline
    Mode                &   {$N_{\rm tag}^{\rm obs, \alpha}$}   &   $\epsilon_{\rm tag}^{\alpha}$(\%) &   $\epsilon_{\rm tag,sig}^{\alpha}$(\%)    \\
    \hline
    $K^+\pi^-$          & 541541(753)    & $65.92\pm0.02$        & $15.18\pm0.20$    \\
    $K^+\pi^-\pi^0$     & 1040340(1209)  & $34.66\pm0.01$        & $8.00\pm0.08$     \\
    $K^+\pi^-\pi^+\pi^-$& 706179(982)    & $38.96\pm0.01$        & $7.02\pm0.09$     \\
    \hline
    $K^+\pi^-\pi^-$     & 806444(953)    & $51.08\pm0.02$        & $5.23\pm0.07$     \\
    $K^+\pi^-\pi^-\pi^0$& 252088(816)    & $25.91\pm0.02$        & $2.40\pm0.06$     \\
    $K_S^0\pi^-$        & 100019(337)    & $54.33\pm0.05$        & $5.55\pm0.21$     \\
    $K_S^0\pi^-\pi^0$   & 235011(759)    & $29.63\pm0.03$        & $3.10\pm0.08$     \\
    $K_S^0\pi^+\pi^-\pi^-$& 131815(710)  & $32.49\pm0.05$        & $2.66\pm0.10$     \\
    $K^+K^-\pi^-$       & 69642(398)     & $40.58\pm0.06$        & $4.09\pm0.20$     \\
    \hline\hline
\end{tabular}
}
\end{footnotesize}
\end{center}
\end{table}

For the semileptonic signal candidate, the undetected neutrino is
inferred by studying the variable $U \equiv E_{\rm miss} - c|\vec{p}_{\rm miss}|$,
where $E_{\rm miss}$ and $\vec{p}_{\rm miss}$ are the missing energy
and momentum carried by the neutrino from the semileptonic decay.  These are
calculated as $E_{\rm miss} = E_{\rm beam} - E_{a_0(980)} - E_{e}$ and
$\vec{p}_{\rm miss} = -(\vec{p}_{\rm tag} + \vec{p}_{a_0(980)} + \vec{p}_{e})$,
respectively,  where $E_{a_0(980)}~$($E_{e}$) and $\vec{p}_{a_0(980)~}$($\vec{p}_{e}$)
are the energy and momentum of $a_0(980)~$(positron), $\vec{p}_{\rm tag}$
is the momentum of the ST $\bar{D}$ in the center-of-mass frame.  We calculate
$\vec{p}_{\rm tag} = \hat{p}_{\rm tag}\sqrt{E_{\rm beam}^2/c^2 - M_D^2c^2}$,
where $\hat{p}_{\rm tag}$ is the unit vector in the momentum direction of the ST $\bar{D}$
and $M_D$ is the nominal $D$ mass~\cite{pdg:2016}. The signal candidates
are expected to peak around zero in the $U$ distribution and near the
$a_0(980)$ mass in the $M_{\eta\pi}$ distribution.

\begin{figure}
\centering
\includegraphics[width=0.48\textwidth]{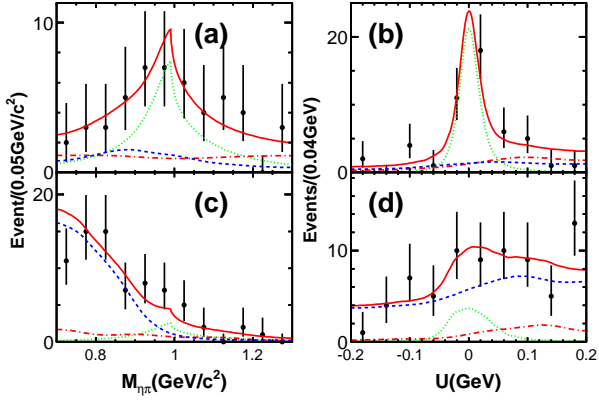}
\caption{(Color online). Projections of the 2-D fit on (left)~$M_{\eta\pi}$ and (right)~$U$ for
    (a)(b)~$\dochannel~$ and (c)(d)~$\dpchannel$. Points with error bars are data.
    The (red) solid curves are the overall fits, the (blue) dashed line
    denote the sum of the Bkg I and Bkg II,  the (red) dotted-dashed lines denote the Bkg III,
    and the (green) dotted lines show the fitted signal shape. }
    \label{fig:fitsig}
\end{figure}

To obtain the signal yields,  we perform two-dimensional~(2-D) unbinned
maximum likelihood fits to the $M_{\eta\pi}$ versus $U$ distributions,
combining all tag modes. Projections of the 2-D fits are shown in Fig.~\ref{fig:fitsig}.
The signal shape in the $U$ distribution is described by the MC simulation
and that in the $M_{\eta\pi}$ distribution is modeled with a usual
Flatt\'e formula~\cite{flatte} for the $a_0(980)$ signal.
The mass and two coupling constants, $g_{\eta\pi}^2$ and $g_{K\bar{K}}^2$
are fixed to $0.990~{\rm GeV}/c^2$, $0.341~({\rm GeV}/c^2)^2~$ and
$0.304~({\rm GeV}/c^2)^2~$\cite{BesIII:2017}, respectively. The backgrounds
are divided into three classes: the residual background from certain specific
$D$ decay modes mentioned previously (Bkg I), the other $D$ decay background
(Bkg II), and the non-$D\bar{D}$ background (Bkg III). For each background
source in Bkg I, the shape and yield are determined by the MC simulation
incorporating the corresponding branching fraction~\cite{pdg:2016}.  The shape and yield
for Bkg II are fixed based on the generic $D\bar{D}$ MC sample, in which
all particles decay inclusively based on the branching fractions taken from
the PDG~\cite{pdg:2016}, but with Bkg I modes removed.
Bkg III, from the continuum processes
$e^+e^- \to q\bar{q}$ and $\tau^+ \tau^-$, is modeled with a MC-determined
shape generated with a modified {\sc lund} model~\cite{lundcharm},
with the yield determined in the fit. The 2-D probability density functions~(PDFs)
of all these components are constructed by the product of the $U$ and $M_{\eta\pi}$
distributions due to the negligible correlation between the two observables
according to the exclusive background channel MC simulation.

The 2-D fits yield $\doresultnsig$ signal events for $\dochannel~$ and $\dpresultnsig$
signal events for $\dpchannel$. The statistical significance of signal, taken
to be $\sqrt{-2\ln({\cal L}_0/{\cal L}_{\rm best})}$, where ${\cal L}_{\rm best}$ and ${\cal L}_0$ are
the maximum likelihood values with the signal yield left free and fixed at zero,
respectively, is $\dosig~$ for $\dochannel$ and $\dpsig~$ for $\dpchannel$.
The corresponding DT efficiencies are presented in Table~\ref{tab:tagEfficiency}.

The systematic uncertainties in the measurements are summarized in
Table~\ref{tab:susum} and discussed below. The uncertainty due to the
ST $\bar{D}$ meson largely cancel in the DT analysis method. The
uncertainties associated with the tracking and PID for the charged
pion are estimated to be $1.0\%$ and $0.5\%$, respectively, by
investigating a control sample $D^+ \to K^-\pi^+\pi^+$ based on
a partial reconstruction technique. Similarly, the uncertainty
related with the $\pi^0$ reconstruction, including the detection
of two photons, is found to be $1.0\%$ by studying the control sample
$D^0 \to K^- \pi^+ \pi^0$. Since $\eta$ candidates are reconstructed
similarly, the corresponding uncertainty is also assigned to be $1.0\%$.
The uncertainties related to tracking and PID for the positron
are investigated with a radiative Bhabha control sample
in the different polar angle and momentum bins. The values for the
tracking and PID are $1.0\%$ and $0.6\%$, respectively, obtained after
re-weighting according to the distributions of momentum and
polar angle of the positron from the signal MC sample. The uncertainty arising
from the choice of the best $\eta\pi^0$ combination in the $D^+$ decay
is studied with a DT $D$ hadronic decay sample,
$D^0 \to K^- \pi^+ \pi^0$ versus $\bar{D}^0 \to K^+ \pi^- \pi^0$
and is taken as $0.3\%$~\cite{combo}. The systematic uncertainty from selecting the best $\eta\pi^0$ combination
is assumed to be the same as the one from selecting the best $\pi^0\pi^0$
combination in Ref.~\cite{combo}, considering
the similar selection criteria of $\eta$ and $\pi^0$.
The efficiency of the lateral moment requirement for photons is studied in
different energy and polar angle bins using a control sample of radiative
Bhabha events.  The average data-MC efficiency difference, after re-weighting
according to the energy and polar angle distributions of the signal MC sample,
is taken as the systematic uncertainty. The form factor of the semileptonic
decay for the nominal signal MC sample is parameterized with the model
of Ref.~\cite{slbkpole}. An alternative MC sample based on
the {\sc isgw2} model~\cite{isgw2} is produced to estimate
the uncertainty associated with the signal model;
the change in the detection efficiency is assigned as the corresponding
systematic uncertainty. The uncertainties in the branching fractions
of submodes are taken from the current world averages~\cite{pdg:2016}.
The effect of limited MC statistics is also included as a systematic effect.
Uncertainties associated with the 2-D fits are estimated by varying
the signal and background shapes and certain background contributions in
Bkg I and Bkg II within their uncertainties. For the resolution of $U$, the
distribution in $U$ of the $D^0$ decay is convolved with a Gaussian
function with free parameters and the fit is redone.
Considering the limited statistics and large background contributions,
the width of the Gaussian function for the $D^+$ decay is fixed to be
$\frac{\rm FWHM_{+}}{\rm FWHM_{0}}\cdot{\sigma_{0}}$,
in which $\sigma_{0}$ is the output Gaussian width in the fit to the $D^0$ case,
and $\rm FWHM_{+}$ and $\rm FWHM_{0}$ are the full width at half maximum of
the nominal $U$ shape for the $D^+$ and $D^0$ signal MC samples,
respectively. Changes in the
signal yields are assigned to be the corresponding uncertainties.
For the $a_0(980)$ line shape, the mass and the two coupling constants in the
Flatt\'e formula are varied by one standard deviation, and the average change
in the signal yield is taken to be the relevant uncertainty. The shapes of the
$D\bar{D}$ and non-$D\bar{D}$ backgrounds are modeled using the Kernel PDF
estimator~\cite{keys} based on the MC samples with a smoothing parameter set to
$1.5$. The uncertainties of the shapes are determined by changing the smoothing
parameter by $\pm0.5$ and we take the relative changes on the signal yield
as the associated uncertainties. We also shift the yields of Bkg I and Bkg II
in the fits by $1\sigma$, calculated from the corresponding branching fractions,
luminosity measurements~\cite{BesLum} and $D\bar{D}$ cross section~\cite{cleo:2014}.
The average changes on the  signal yields are taken as the corresponding
uncertainties.

\begin{table}[hbtp]
\caption{{\small{The relative systematic uncertainties (in \%) on the branching
    fraction measurements. Items marked with $*$ are derived from the
    fit procedure and are not used when evaluating the upper
    limit of the branching fraction.}}}\label{tab:susum}
\begin{center}
\begin{footnotesize}
\renewcommand{\arraystretch}{1.0}
\scalebox{0.95}{
  \begin{tabular}{ccc}
    \hline \hline
    Source                 &   $D^0 \to a_0(980)^- e^+ \nu_e$          & $D^+ \to a_0(980)^0 e^+ \nu_e$     \\
    \hline
    Tracking               &         $2.0$                             &       $1.0$                        \\
    $\pi$ PID              &         $0.5$                             &           -                        \\
    $\pi^0$ reconstruction &            -                              &       $1.0$                        \\
    $\eta$ reconstruction  &         $1.0$                             &       $1.0$                        \\
    Positron PID           &         $0.6$                             &       $0.6$                        \\
    The best $\eta \pi^0$ combination  &     -                         &       $0.3$                        \\
    Lateral moment requirement     &         $1.6$                     &       $1.6$                        \\
    Form factor model           &         $5.3$                             &       $5.6$                        \\
    $\eta$ and $\pi^0$ branching fraction &  $0.5$                     &       $0.5$                        \\
    MC statistics          &         $0.6$                             &       $0.9$                        \\
    *$U$ resolution        &         $2.7$                             &       $1.1$                       \\
    *$a_0(980)$ line shape &         $0.2$                             &       $0.3$                       \\
    *Background modeling   &         $0.3$                             &       $2.0$                       \\
    \hline
    Total                  &         $6.7$                             &       $6.6$                        \\
    \hline \hline
  \end{tabular}
}
\end{footnotesize}
\end{center}
\end{table}

Due to the limited statistical significance of the $\dpchannel$ mode,
an upper limit on the signal yield is also computed using a Bayesian method. The fit likelihood as
a function of the number of signal events, denoted as $f_{\cal L}(N)$,
is convolved  with Gaussian functions that represent the systematic uncertainties.
For all uncertainty sources except those from the 2-D fit, the effects are taken into account by
Gaussian functions having widths equal to the corresponding uncertainties.
Uncertainties due to the fit procedure are computed by varying choices
of fit conditions to toy MC simulated events, sampled according to the shape of data. In
each toy experiment, we perform a nominal fit and one alternative fit
with the shape parameters varied in the fit procedure as described above.
A Gaussian function is obtained with parameters taken from the mean and the
root-mean-square of the resultant discrepancy between the two fitted yields.
By integrating up to $90\%$ of the physical region for the smeared $f_{\cal L}(N)$, we
obtain an upper limit of $N^{\rm up}<\dpupnsig$ at the $90\%$ confidence level (C.L.)
for the $\dpchannel$ yield.

Since the branching fraction of $a_0(980) \to \eta \pi$ has not been
well-measured, we report the product branching fractions, obtaining
\begin{equation}
\nonumber
{
\begin{aligned}
    \mathcal{B}&(\dochannel)\times\mathcal{B}(\aomdecay)\\
    &= \doresult\\
\end{aligned}
}
\end{equation}

\begin{equation}
\nonumber
{
\begin{aligned}
    \mathcal{B}&(\dpchannel)\times\mathcal{B}(\aoodecay)\\
    &= \dpresult,\\
\end{aligned}
}
\end{equation}
where the first (second) uncertainties are statistical (systematic).
The upper limit on the product branching fraction for $D^+$ decay is determined as
$\mathcal{B}(\dpchannel)\times\mathcal{B}(\aoodecay)\dpul$ at the $90\%$ C.L.
By convolving the likelihood value from the nominal fits with Gaussian
functions whose widths represent the systematic uncertainties for the $D^0$
and $D^+$ decays, we calculate the signal significance including systematic
uncertainties to be $\dosigs$ and $\dpsigs$ for the $D^0$ and $D^+$ decays,
respectively.

To summarize, we present the observation of the semileptonic decay of
$\dochannel$ and the evidence for $\dpchannel$. Taking the lifetimes
of $D^0$ and $D^+$~\cite{pdg:2016} into consideration and
assuming that $\mathcal{B}(\aomdecay) = \mathcal{B}(\aoodecay)$,
we find a ratio of partial widths of
\begin{equation}
\nonumber
    \frac{\Gamma(\dochannel)}{\Gamma(\dpchannel)} = 2.03 \pm 0.95 \pm 0.06 {\rm ,}
\end{equation}
consistent with the prediction of isospin symmetry, where the shared
systematic uncertainties have been canceled.  The two branching
fractions provide information about the $d\bar{u}$
and $(u\bar{u}-d\bar{d})/\sqrt{2}$ components in the $a_0(980)^-$
and $a_0(980)^0$ wave functions, respectively~\cite{Achasov:2012}.
Along with the result of the branching fraction of $D^+ \to f_0 e^+ \nu_e$,
a result in preparation at BESIII, we will have a valuable input for
understanding the nature of the light scalar mesons.
\vspace{2.0cm}

The BESIII collaboration thanks the staff of BEPCII and the IHEP computing center for their strong support. This work is supported in part by National Key Basic Research Program of China under Contract No. 2015CB856700; National Natural Science Foundation of China (NSFC) under Contracts Nos. 11235011, 11335008, 11425524, 11625523, 11635010; the Chinese Academy of Sciences (CAS) Large-Scale Scientific Facility Program; the CAS Center for Excellence in Particle Physics (CCEPP); Joint Large-Scale Scientific Facility Funds of the NSFC and CAS under Contracts Nos. U1332201, U1532257, U1532258; CAS under Contracts Nos. KJCX2-YW-N29, KJCX2-YW-N45, QYZDJ-SSW-SLH003; 100 Talents Program of CAS; National 1000 Talents Program of China; INPAC and Shanghai Key Laboratory for Particle Physics and Cosmology; German Research Foundation DFG under Contracts Nos. Collaborative Research Center CRC 1044, FOR 2359; Istituto Nazionale di Fisica Nucleare, Italy; Joint Large-Scale Scientific Facility Funds of the NSFC and CAS; Koninklijke Nederlandse Akademie van Wetenschappen (KNAW) under Contract No. 530-4CDP03; Ministry of Development of Turkey under Contract No. DPT2006K-120470; National Natural Science Foundation of China (NSFC) under Contract No. 11505010; National Science and Technology fund; The Swedish Resarch Council; U. S. Department of Energy under Contracts Nos. DE-FG02-05ER41374, DE-SC-0010118, DE-SC-0010504, DE-SC-0012069; University of Groningen (RuG) and the Helmholtzzentrum fuer Schwerionenforschung GmbH (GSI), Darmstadt; WCU Program of National Research Foundation of Korea under Contract No. R32-2008-000-10155-0.

\input{bibliography.tex}
\end{document}

%% file: authors.tex
\author{
  \begin{small}
    \begin{center}
        M.~Ablikim$^{1}$, M.~N.~Achasov$^{9,d}$, S.~Ahmed$^{14}$, M.~Albrecht$^{4}$, 
        A.~Amoroso$^{53A,53C}$, F.~F.~An$^{1}$, Q.~An$^{50,40}$, J.~Z.~Bai$^{1}$, 
        Y.~Bai$^{39}$, O.~Bakina$^{24}$
% ...
\\
\vspace{0.2cm}
(BESIII Collaboration)\\
\vspace{0.2cm} {\it
$^{1}$ Institute of High Energy Physics, Beijing 100049, People's Republic of China\\
% ...
$^{4}$ Bochum Ruhr-University, D-44780 Bochum, Germany\\
% ...
$^{9}$ G.I. Budker Institute of Nuclear Physics SB RAS (BINP), Novosibirsk 630090, Russia\\
% ...
$^{14}$ Helmholtz Institute Mainz, Johann-Joachim-Becher-Weg 45, D-55099 Mainz, Germany\\
% ...
$^{24}$ Joint Institute for Nuclear Research, 141980 Dubna, Moscow region, Russia\\
% ...
$^{39}$ Southeast University, Nanjing 211100, People's Republic of China\\
$^{40}$ State Key Laboratory of Particle Detection and Electronics, Beijing 100049, Hefei 230026, People's Republic of China\\
% ...
$^{50}$ University of Science and Technology of China, Hefei 230026, People's Republic of China\\
% ...
$^{53}$ (A)University of Turin, I-10125, Turin, Italy; (B)University of Eastern Piedmont, I-15121, Alessandria, Italy; (C)INFN, I-10125, Turin, Italy\\
% ...
$^{d}$ Also at the Novosibirsk State University, Novosibirsk, 630090, Russia\\
% ...
}
\end{center}
\vspace{0.4cm}
\end{small}
}

%% file: a_DtoA0enu.bbl
\begin{thebibliography}{**}
\bibitem{Amsler:2013}
    C.~Amsler {\it et al.}, 
    ``Note on Scalar Mesons below 2 GeV'',
    Review published on Particle Data Group,
    Chin.\ Phys.\ C {\bf 40}, 100001 (2016).


\bibitem{Oset:2015}
    T.~Sekihara and E.~Oset,
    Phys.\ Rev.\ D {\bf 92}, 054038 (2015).

\bibitem{Xiao-Dong:2017}
    X.~D.~Cheng {\it et al.}, 
    Phys.\ Rev.\ D {\bf 96}, 033002 (2017).

\bibitem{Achasov:2012}
    N.~N.~Achasov and A.~V.~Kiselev,
    Phys.\ Rev.\ D {\bf 86}, 114010 (2012).

\bibitem{Ke:2009}
    H.~W.~Ke, X.~Q.~Li and Z.~T.~Wei,
    Phys.\ Rev.\ D {\bf 80}, 074030 (2009).

\bibitem{Bennich:2009}
    B.~El-Bennich, O.~Leitner, J.-P.~Dedonder and B.~Loiseau,
    Phys.\ Rev.\ D {\bf 79}, 076004 (2009).

\bibitem{Lv:2010}
    W.~Wang and C.~D.~L\"u,
    Phys.\ Rev.\ D {\bf 82}, 034016 (2010).

\bibitem{BesLum}
    M.~Ablikim {\it et al.}  [BESIII Collaboration],
    Chin.\ Phys.\ C {\bf 37}, 123001 (2013);
    Phys.\ Lett.\ B {\bf 753}, 629 (2016).

\bibitem{bes3}
    M.~Ablikim {\it et al.}  [BESIII Collaboration],
    Nucl.\ Instrum.\ Meth.\ A {\bf 614}, 345 (2010).

\bibitem{geant4}
  S.~Agostinelli {\it et al.}  [GEANT4 Collaboration],
  Nucl.\ Instrum.\ Meth.\ A {\bf 506}, 250 (2003).

\bibitem{kkmc}
  S.~Jadach, B.~F.~L.~Ward and Z.~Was,
  Comput.\ Phys.\ Commun.\  {\bf 130}, 260 (2000);
  Phys.\ Rev.\ D {\bf 63}, 113009 (2001).

\bibitem{evtgen}
  D.~J.~Lange,
  Nucl.\ Instrum.\ Meth.\ A {\bf 462}, 152 (2001);
   R.~G.~Ping,
  Chin.\ Phys.\ C {\bf 32}, 599 (2008).

\bibitem{photons}
  E.~Richter-Was,
  Phys.\ Lett.\ B {\bf 303}, 163 (1993).

\bibitem{MarkIII:1986}
  R.~M.~Baltrusaitis {\it et al.}  [MARK III Collaboration],
  Phys.\ Rev.\ Lett.\  {\bf 56}, 2140 (1986).

\bibitem{ycp}
  M.~Ablikim {\it et al.}  [BESIII Collaboration],
  Phys.\ Lett.\ B {\bf 744}, 339 (2015).

\bibitem{d0semilep}
  M.~Ablikim {\it et al.}  [BESIII Collaboration],
  Phys.\ Rev.\ D {\bf 92}, 072012 (2015).  

\bibitem{He:2005bs}
  Q.~He {\it et al.}  [CLEO Collaboration],
  Phys.\ Rev.\ Lett.\  {\bf 95}, 121801 (2005).

\bibitem{Albrecht:1990am}
  H.~Albrecht {\it et al.}  [ARGUS Collaboration],
  Phys.\ Lett.\ B {\bf 241}, 278 (1990).

\bibitem{pdg:2016}
    C.~Patrignani {\it et al.}~[Particle Data Group],
    Chin.\ Phys.\ C {\bf 40}, 100001 (2016).

\bibitem{latmom}
    A.~Drescher {\it et al.}, 
    Nucl.\ Instrum.\ Meth.\ A {\bf 237}, 464 (1985).

\bibitem{flatte}
    S.~M.~Flatt\'e,
    Phys.\ Lett.\ B {\bf 63}, 224 (1976).

\bibitem{BesIII:2017}
    M.~Ablikim {\it et al.}  [BESIII Collaboration],
    Phys.\ Rev.\ D {\bf 95}, 032002 (2017).

\bibitem{lundcharm}
    J.~C.~Chen, G.~S.~Huang, X.~R.~Qi, D.~H.~Zhang and Y.~S.~Zhu,
    Phys.\ Rev.\ D {\bf 62}, 034003 (2000).

\bibitem{combo}
    M.~Ablikim {\it et al.}  [BESIII Collaboration],
    Chin.\ Phys.\ C {\bf 40}, 113001 (2016).

\bibitem{slbkpole}
    D.~Becirevic and A.~B.~Kaidalov,
    Phys.\ Lett.\ B {\bf 478}, 417 (2000).

\bibitem{isgw2}
    D.~Scora and N.~Isgur,
    Phys.\ Rev.\ D {\bf 52}, 2783 (1995).

\bibitem{keys}
    K.~S.~Cranmer, 
    Comput.\ Phys.\ Commun., {\bf 136}, 198 (2001).


\bibitem{cleo:2014}
    G.~Bonvicini~ {\it et al.} [CLEO Collaboration],
    Phys.\ Rev.\ D {\bf 89}, 072002 (2014).

\end{thebibliography}
